\documentclass[aps,prd,nofootinbib,superscriptaddress,twocolumn]{revtex4-1}
\usepackage{latexsym}
\usepackage{amsthm}
\usepackage{amsmath}
\usepackage{graphicx}
\usepackage{amssymb}
\usepackage{hyperref}
\usepackage{bbm}
\newcommand{\gsim}{\lower.7ex\hbox{$\;\stackrel{\textstyle>}{\sim}\;$}}
\newcommand{\lsim}{\lower.7ex\hbox{$\;\stackrel{\textstyle<}{\sim}\;$}}
\def\ee{\end{eqnarray}}

\def\nn{\nonumber}

\newcommand{\bea}{\begin{eqnarray}}
\newcommand{\eea}{\end{eqnarray}}

\newcommand{\comment}[1]{}

\begin{document}

\pagestyle{plain}

\title{Electrons turn into anyons under an elastic membrane}

\author{Siavosh R. Behbahani}
\author{Claudio Chamon}
\author{Emanuel Katz}
\affiliation{
Physics Department, 
Boston University, 
Boston, Massachusetts 02215, USA
            }
\date{\today}
\begin{abstract}
  We show that electrons acquire anyonic statistics if a
  two-dimensional electron gas is placed in proximity to a
  charged magnetic membrane. In this two-layered system,
  electrons become anyons through electromagnetic field mediated
  interactions with a membrane phonon. The statistical parameter is
  irrational and depends on the membrane tension and on the electric
  charge and magnetic dipolar densities on the membrane. We propose
  a measurement that yields the value of the statistical parameter.
  
\end{abstract}

\maketitle

\section{introduction}

Anyons are particles whose wavefunctions acquire a nontrivial phase
upon exchange of coordinates, {\it i.e.}, a phase factor other than
that for bosons ($+1$) or fermions
($-1$)~\cite{Leinaas-Myrheim,Wilczek}. Anyons are possible in two
spatial dimensions (2D), where the braiding of the worldlines of the
particles is non-trivial. Quasiparticle excitations that obey Abelian
anyonic exchange statistics, as well as non-Abelian extensions, are
theoretically predicted to exist in fractional quantum Hall (FQH)
states~\cite{Halperin,Arovas}. Much effort, in the form of theoretical
proposals~\cite{CFKSW,FNTW} and experimental
attempts~\cite{Goldman1,Goldman2,Willet,Kang}, has been devoted to the
detection of both Abelian and non-Abelian anyons in FQH liquids.

In the FQH effect, the elementary quasiparticle excitations carry both
rational fractional charges and rational statistics (the phases are
rational multiples of
$\pi$)~\cite{Laughlin,Haldane,Halperin,Arovas}. This ``rationality''
is rooted in the connection of the FQH effect and Chern-Simons
theory~\cite{Read90,Blok90,Frohlich91b,Wen92a,Frohlich94},
from which it also follows that the many-body ground state is
degenerate on closed surfaces, with the degeneracy depending on the
surface genus~\cite{Wen89,Wen1990}.

Anyons, however, need not be always rational like in the FQH
effect. The statistical angles can in principle take irrational
values~\cite{Wilczek}, once the physical origin of the anyonic
statistics is divorced from Chern-Simons theory. One example where the
statistical angle can vary continuously is when the quasiparticles are
associated to vortices in a complex mass order parameter (a Higgs
field) for 2D Dirac electrons~\cite{Gang-of-6,RMHC}. In these examples
both the quasiparticle charge and statistics can vary continuously,
but then the vortices are logarithmically confined. The vortices can
be deconfined with the addition of an axial gauge
field~\cite{Jackiw-Pi}, but when they are, the charge and statistics
become rational again~\cite{Gang-of-6,RMHC}.

The goal of this work is to point out that, given a recently suggested
alternative framework for realizing anyonic statistics~\cite{Katz}, it
is possible to turn ordinary electrons into anyons. This mutation, we
show, occurs when a 2D electron gas (2DEG) is placed under an elastic
electric and magnetic membrane. In this two-layered system, the anyons
reside on the 2DEG layer. The essence of the mechanism is that height
fluctuations of the membrane generate fluctuations in the gauge
potential seen by the electrons in the 2DEG below, while the charge
density of the electrons can affect the height fluctuations of
membrane above. Hence, the membrane fluctuations act as a middle man,
allowing the electron charge to directly source the magnetic
field. This is sufficient to turn an electron into an anyon.

Under the elastic membrane, the electric charge of the particle is
rational (in fact, integer), but the statistical angle is not forced
to take rational values. Instead, it depends continuously on the
elastic membrane tension, $\tau$, as well as on the electric charge
and magnetic dipole densities $(\sigma_e,\sigma_m)$ on the
membrane. The statistical angle is $\pi+\theta$, where
\begin{eqnarray}
\label{eq:theta-statistics}
\theta= -2\pi^2 e^2 ~\frac{\sigma_e\sigma_m}{\tau}
\;.
\end{eqnarray} 
We present the mechanism behind the statistical transmutation of
electrons into anyons under the electric and magnetic membrane, and
argue that there is an experimentally accessible regime where the
effect could be probed.

The basic observation underlying the change in electron statistics is
the fact that in 2+1 dimensions a gauge field has the same degrees of
freedom as a scalar. Thus, the anyonic phenomena can be obtained via
the replacement of the spatial components of a statistical gauge
field, $a_i$, by a scalar field: $a_i = \epsilon_{ij} \nabla_j
\phi$. In terms of the scalar, the statistical magnetic field
$b= \epsilon_{ij} \nabla_i a_j = -\nabla^2\phi$.  We conclude that
anyons will result if we can arrange that the electron field,
$\psi(\vec r)$, serves as a source for $\phi(\vec r)$,
\begin{eqnarray}
\label{gsource}
\nabla^2 \phi(\vec r) &=& \frac{g}{\tau} \; \psi^\dag \psi(\vec r),
\end{eqnarray} 
while at the same time enjoying a coupling to it through a ``covariant
derivative'' which mimics the effects of a gauge field $a_i$:
\begin{eqnarray}
\label{alpha}
{\cal L}= -\frac{1}{2m}
\left|\left(\nabla_i 
- i\alpha\, \epsilon_{ij} \nabla_j \phi \right) \psi \right|^2.
\end{eqnarray}
In this framework, the statistical parameter $\theta$ is given in
terms of the couplings $g/\tau$ and $\alpha$ as
$\theta=\frac{g\alpha}{2\tau}$, and can be any real number.

An action which leads to (\ref{gsource}) (at sufficiently
large distances) and (\ref{alpha}) is
\begin{eqnarray}
S =&& \int\! dt\, d^2x\, \bigg[
\frac{\rho}{2}\, \dot \phi^2 
-\frac{\tau}{2}\, (\nabla \phi)^2  
-\frac{\kappa}{2}\, (\nabla^2 \phi)^2
\nn\\ 
&&\;\;+\;\psi^\dag i\partial_t \psi
-\frac{1}{2m} \left| \left(\nabla_i - i\alpha\, \epsilon_{ij} \nabla_j \phi \right) \psi \right|^2  
\nn\\ 
&&\;\; -\;g\,  \phi\;\,  \psi^\dag\psi\; \bigg]\,.
\label{eq:ourLag}
\end{eqnarray}
The first line describes the height fluctuations of a membrane under
tension $\tau$.  The $\kappa$ term is the extrinsic curvature
contribution to the membrane bending energy, and $\rho$ is the surface
mass density. Our task in what follows will be to induce the couplings
on the second and third line above between an electron and the
membrane phonon. We will see that this is possible through the
electromagnetic fields due to electric charges and magnetically
aligned dipoles on the elastic membrane.

\section{Potentials under the membrane}

\begin{figure}[setup]
\scalebox{0.25}{\includegraphics{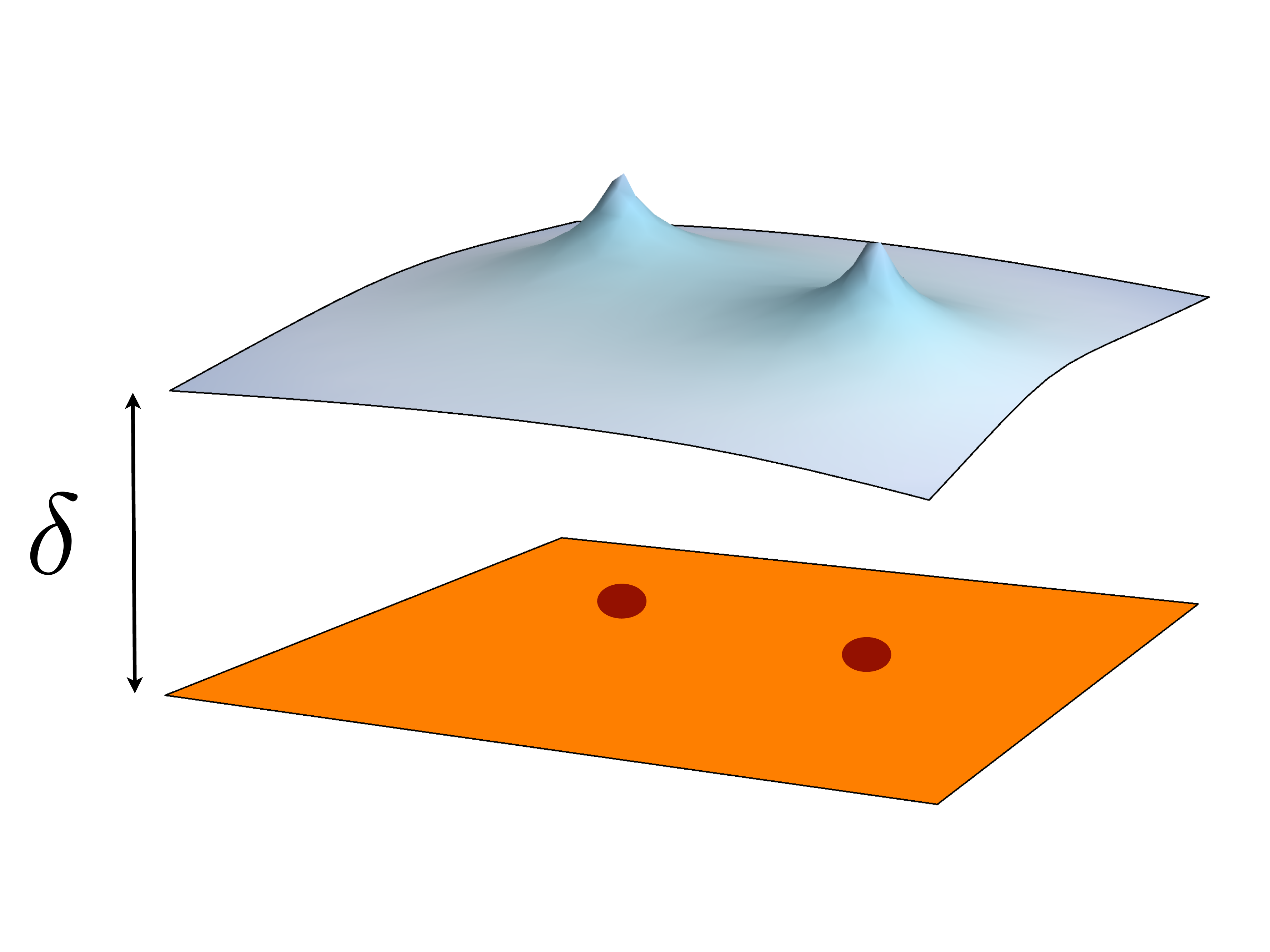}}
\caption{\label{fig:typical}This Figure illustrates the elastic
  electric and magnetic membrane on top of the metallic sheet a
  distance $\delta$ away. Red spots are schematically showing the
  electrons on the metallic sheet. The presence of the electrons
  deflect the charged membrane, in turn generating a magnetic gauge
  potential that is sensed by other electrons on the plane.}
\label{fig:setup}
\end{figure}

Consider a magnetic membrane, with an areal density $\sigma_m$ of
magnetic dipolar moments that are aligned along the normal to the
membrane. The orientation of the dipoles can be fixed by a
perpendicular external magnetic field. We note that because of the
external field, one can equally use a paramagnetic or ferromagnetic
system.   In addition, let the membrane posses a charge density, $\sigma_e$.
Also consider a metallic sheet, with some electronic density
a distance $\delta$ away from the membrane. For
simplicity we assume that the separation $\delta$ is much smaller than
the length of the membrane.
The configuration is depicted in Fig.~\ref{fig:setup}.

We assume the following Hamiltonian for electrons on the metallic sheet:
\begin{eqnarray}
\label{eq:e-H}
H_{\psi} &=& \frac{1}{2m}\left(\vec{p}+e\vec{A}\right)^2\, - e A_0 \,\,.
\end{eqnarray}
Here $A_0$ is the electric potential, and $A_i$ is the magnetic vector
potential.

Our first task is to calculate the vector potential on the metallic
sheet induced by the height fluctuations of the membrane. In this
calculation we assume that the surface of the magnetic membrane
fluctuates while the dipolar moments are fixed in the $\hat{z}$
direction. This assumption is reasonable for two reasons. The first is
that corrections due to the tilting of the dipoles is of higher order
in gradients of $\phi$. The second is that the tilting of the dipoles
is gapped by the external field that aligns the spins along the $\hat
z$ direction.

We assign the surface $X'=(x',y')$ coordinate to the membrane and
$X=(x,y)$ to the metallic sheet. The vector potential on the metallic
sheet is calculated from
\begin{eqnarray}
A_i(X)\!=\!\epsilon_{ij}\!\!\!\int \!d^2 X'
\frac{-\tilde\sigma_m(X')\; (X-X')_j}{\left[\left(X-X'\right)^2+\left(\delta + \phi\left(X'\right)\right)^2\right]^{3/2}},
\end{eqnarray}
where $\phi(X')$ is the height fluctuation of the membrane around
its average position $\delta$. (Notice that $\delta>0$ if the membrane
is placed above the metallic sheet, and $\delta<0$ when the membrane
is below.) 
We also introduced
\begin{eqnarray}
\tilde\sigma_m(X')=\frac{\sigma_m}{\sqrt{1+\left(\nabla\phi(X')\right)^2}},
\end{eqnarray}
which is related to $\sigma_m$ by considering the tilting of the
normal direction. 
In the absence of height fluctuations, the magnetic
membrane induces a constant magnetic field on the metallic
sheet. Therefore we can decompose the vector potential as
\begin{eqnarray}
\vec{A}=\vec{A}^{BG}+\vec{A^\phi}
\;.
\end{eqnarray}
The contribution $\vec{A}^{BG}$ leads to a uniform background field,
which can be added to the externally applied field. The contribution
$\vec{A^\phi}$ is due to the height fluctuations of the membrane. We
assume that these fluctuations are smooth on the scale of $\delta$. 
As we show in the Appendix, the vector potential fluctuation is
\begin{eqnarray}
    A^\phi_i(X)&\approx& 2\pi\,\sigma_m\,\epsilon_{ij}\nabla_j \phi(X)\,{\rm sgn}(\delta).
\label{eq:A_i}
\end{eqnarray}
Notice that to leading order the vector potential is independent of
the magnitude of $\delta$, but it depends on its sign, {\it i.e.}, on
whether the membrane placed above or below the electronic
plane. 

A very similar calculation leads to the dependence of the electric potential
on the membrane fluctuation:  
\begin{eqnarray}
A_0^\phi(X)&\approx& -2\pi\,\sigma_e\,\phi(X)\,{\rm sgn}(\delta).
\end{eqnarray}
Here, as above, we have ignored the uniform background electric
field in the $\hat z$ direction.

Comparing (\ref {eq:ourLag}) with (\ref{eq:e-H}) we can read off the
couplings of the membrane phonons with the electrons of the metallic
layer:
\begin{eqnarray}
&&
\alpha =\, -2\pi\,\sigma_m\;e\;{\rm sgn}(\delta)\;,
\quad
g=2 \pi\,\sigma_e\;e\;{\rm sgn}(\delta)
\;.
\end{eqnarray} 

Thus, in terms of the membrane parameters, the statistical angle
$\theta=\frac{g\alpha}{2\tau}$ is given by the expression presented in
Eq.~(\ref{eq:theta-statistics}).  We note that $\theta$ can be any
real number, and can be varied experimentally. Next, we discuss an
experimental setup to measure the statistical parameter $\theta$.

\section{Proposed experimental setup}

\begin{figure}
  \centering
  \scalebox{0.23}{\includegraphics{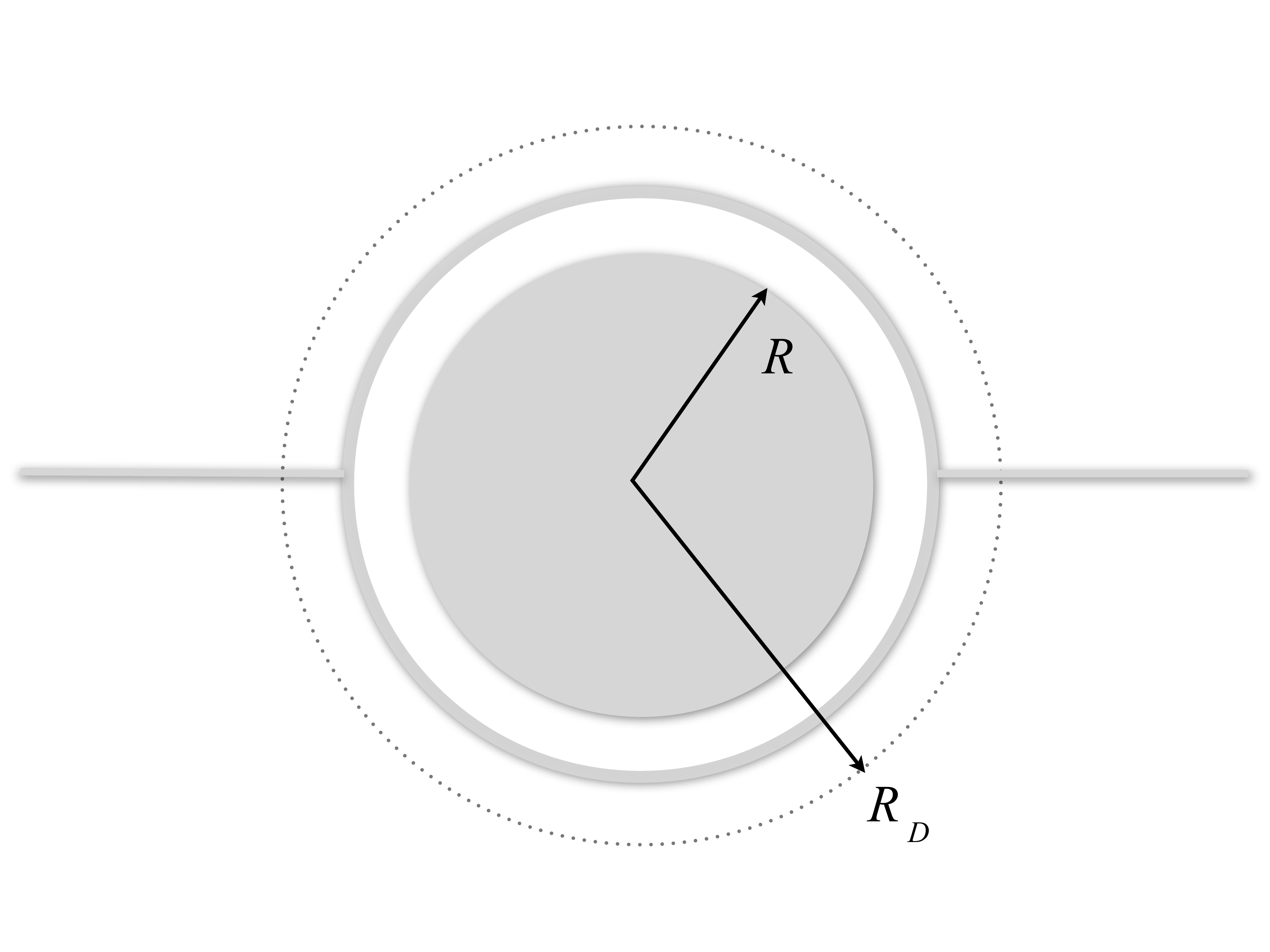}}
  \caption{Geometry to detect the quantal phases acquired by electrons
    due to the membrane above them. The geometry is to be patterned on
    the 2DEG layer underneath the elastic membrane. Electrons loaded
    into the metallic island of radius $R$ cause the membrane to
    displace vertically. The membrane can be clamped down at a
    distance $R_D$ from the center (but not on the 2DEG plane). The
    membrane height displacements generate a magnetic flux for the
    electrons in the metallic ring. The flux gives rise to
    Aharonov-Bohm oscillations of the conductance, which can be
    measured experimentally. For the electrons confined to their
    plane, unaware of the membrane in the third dimension, the
    accumulated quantum mechanical phase is due to anyonic
    statistics.}
  \label{fig:AB-ring-island}
\end{figure}

Consider the geometry depicted in Fig.~\ref{fig:AB-ring-island}, where
a metallic ring and a central metallic island are defined on the 2DEG
plane. The island could be charged by a back gate. The proposed
measurement consists of studying the oscillations of the conductance
through the ring as function of the charge in the island. These
oscillations arise because the charge in the island vertically
displaces the membrane, which in turn generates a magnetic flux for
the electrons in the ring. For the electrons confined to the plane of
the 2DEG, unaware of the membrane in the third dimension, the
accumulated quantum mechanical phase is due to anyonic statistics.

Let us compute the Aharonov-Bohm (AB) phase accumulated by the
electrons in the ring when one charges up the island with charge
$Q$. The deformation of the membrane follows from the equations of
motion for the displacement $\phi$:
\begin{eqnarray}
\label{eq:mech-elec}
\nabla^2\phi(\vec r) = \frac{g}{\tau}\, \rho(\vec r)
-
\frac{\alpha}{\tau}\, \hat z \cdot \,\vec\nabla\times \vec J(\vec r)
\;,
\end{eqnarray}
where $\rho(\vec r)$ and $\vec J(\vec r)$ are, respectively, the
electron number and current densities on the 2DEG plane. Let us focus
on the case where the charge in the island is static ($\vec
J=0$). Because the island is metallic, the charge will concentrate at
its edges. To compute the magnetic flux through a region encircling
the island, due to the charge $Q$ within, we use that (\ref{eq:A_i})
implies that $B=\frac{\alpha}{e}\,\nabla^2\phi$, which gives
\begin{eqnarray}
\label{eq:flux}
\Phi=-\frac{\alpha g}{\tau}\,\frac{Q}{e^2}
\;.
\end{eqnarray}
This result means that flux $\Phi$ is affixed to the charge $Q$. The
particles are a meld of charge and flux, and it is this composite
nature that gives rise to the anyonic statistics.

The flux in~(\ref{eq:flux}) is in addition to any other flux
$\Phi_{\rm ext}$ due to externally applied magnetic fields. The
conductance through the ring, $G(\Phi_{\rm total}/\Phi_0)$, oscillates
as function of $\Phi_{\rm total}=\Phi+\Phi_{\rm ext}$ with period
$\Phi_0=hc/e$ ($=2\pi/e$ in natural units where $\hbar=c=1$). Now, the
charge in the island is quantized, $Q=n\,e$, $n\in \mathbb Z$. Thus
$\Phi$ can only change in steps as the charge is varied by changing
the back gate ($\Delta Q=C\, \Delta V$, where $C$ is the capacitance
of the island). Every time extra $n=-\Delta Q/e$ electrons are added to
the island, the pattern of AB oscillations in the conductance $G$ is
shifted by a phase $2\pi\,\Delta\Phi/\Phi_0=-({\alpha g}/{\tau})\,
\Delta Q/e$. So measurements for different values of back gate
voltages should yield a family of phase shifted patterns of AB
oscillations in the conductance $G$ as function of the externally
applied flux $\Phi_{\rm ext}$. The shift in the patterns is
$\Delta\Phi_{\rm ext}=-\Delta\Phi$.

The statistical parameter $\theta$ of the anyons can then be
determined from measurements of $\Delta\Phi_{\rm ext}$ and $\Delta V$:
\begin{eqnarray}
\label{eq:theta}
{\theta}=\frac{\alpha g}{2\tau}=
\pi\,\frac{\Delta\Phi_{\rm ext}/\Phi_0}{\Delta Q/e}
=\frac{1}{2}\,\frac{e^2}{C^{\,}}\; \frac{\Delta\Phi_{\rm ext}}{\Delta V}
\;.
\end{eqnarray}

A comment is in order. The coupling of the electrons in the 2DEG with
the phonon mode in the membrane leads to an effective
logarithmic potential between the electrons.  A possible worry is
then that a dynamical phase accumulates as the electrons go around the
ring.  Such a phase would depend on the detailed shape of the
ring. The reason why this additional dynamical phase is not present is
that the ring is metallic, and therefore charges can redistribute so
as to bring the total potential (the sum of effective plus electric
potentials) to a constant value throughout the ring. As long as the
width of the ring is smaller than the distance to the membrane,
these redistributed charges will not affect the membrane
displacement above the ring.  Consequently, the electromagnetic
field generated by the membrane will not be affected, which means
that the AB-phase will not be disturbed.
 
\subsection{Possible realizations}

Let us discuss possible examples of such two-layered
magnetic/electronic system. There are many possibilities for the 2DEG,
for example Si or a semiconductor GaAs/AlGaAs heterostructure. The
ability to gate the electron gas, so as to define an Aharonov-Bohm
ring and an island, is important. Alternatively, one could just
deposit thin metallic (say gold) layers on a substrate, and write
directly the wire and island with metal. Essential is the ability to
back gate the island, to control the electron density in it.

As for the possible realization of the electric and magnetic membrane,
perhaps the ideal candidate would be a suspended graphene sheet that
is doped with magnetic ions. The polarized membrane would result from
an externally applied field. By electrically biasing the membrane, one
could charge it too. It would thus be possible to control both the
charge and magnetic dipole densities $\sigma_e$ and $\sigma_m$.

Let us first estimate the size of the various effects we have
discussed in this particular realization. We will only provide an
order of magnitude estimate. Magnetic impurities spaced about $1~nm$
away, with each impurity having a magnetic moment of order $\mu_B$,
yields a magnetic density of order $\sigma_m \sim 3 \times 10^{-3}
\,eV$. 
As for the membrane tension and surface charge density, it is more
suiting to work with the membrane displacements directly. We shall
work with the constraint that the maximum displacement $\phi_{\rm
  max}$ does not exceed the order of $300~nm$. (The distance $\delta$
can be taken to be of order of $\phi_{\rm max}$ safely if the membrane
is poped upwards.)

If the membrane is clamped (Dirichlet boundary conditions) at a radius
$R_D$ from the center of the island (see
Fig.~\ref{fig:AB-ring-island}), then the displacement due to a charge
$Q$ in the island of radius $R$ is
\begin{eqnarray}
\phi(\vec r)=
\begin{cases}
0, & R_D<r \\
-\frac{g}{2\pi\tau} \frac{Q}{e}\,\ln R_D/r, & R<r\le R_D\\
-\frac{g}{2\pi\tau} \frac{Q}{e}\,\ln R_D/R, & r\le R
\;.
\end{cases}  
\end{eqnarray}
We shall consider $R\sim 3\mu m$ and $\Delta R= R_D-R \sim 0.1\,R \sim
300~nm$. The AB ring width must fit within this gap, so a width of
$\sim 30 ~nm$ is reasonable.

One can express the statistical phase $\theta$ in terms of the maximum
displacement of the membrane, at the edge of the island:
\begin{eqnarray}
\label{eq:theta_height}
\frac{\theta}{\pi}=\frac{\alpha g}{2\pi\tau}
&=&
-\frac{\alpha\,\phi_{\rm max}}{Q/e}
\frac{1}{\ln R_D/R}
\nonumber\\
&=&
-\frac{2\pi \sigma_m \,e\,\phi_{\rm max}}{Q/e}
\frac{1}{\ln R_D/R}
\;.
\end{eqnarray}
The statistical phase an electron in the ring feels as a result of all
the $-Q/e$ electrons in the island is
\begin{eqnarray}
-\frac{Q}{e}\frac{\theta}{\pi}
&=&
2\pi \sigma_m \,e\,\phi_{\rm max}
\,\frac{1}{\ln R_D/R}
\nonumber\\
&\approx&
2\pi \sigma_m \,e\,\phi_{\rm max}
\,\frac{R}{\Delta R}
\nonumber\\
&\sim&
3\times 10^{-2}
\;.
\end{eqnarray}
We remark that using ions with magnetic moment larger that one Bohr
magneton and increasing the density of magnetic dopants could raise
this number by one or two orders of magnitude. Furthermore, a larger
maximum displacement $\phi_{\rm max}$ and larger radius $R$
(keeping constant the difference $\Delta R$) can increase the effect by
perhaps another order of magnitude. 

Finally, the value of $Q=-eN_e$ will ultimately depend on the tension
and electric charge density. Because the bending rigidity in graphene
is $\kappa\sim 1~eV$, there is a characteristic scale $L_\tau^2 =
{\kappa}/{\tau}$ which needs to be kept smaller than the dimensions in
the proposed experimental set up (say $L_\tau\sim 10nm$). We can
express the membrane displacement as $\phi_{\rm max} \sim {\sigma_e e
  L_\tau^2 N_e}/{\kappa}$. Parameterizing the charge density as
$\sigma_e = \frac{e}{d^2}$, with $d\sim 1nm$ one would have a charge
of about $10^{-2} e$ per atom in the membrane. We thus find that
$N_e\sim 1$. Therefore the phase picked by the electrons around the AB
ring is sizable even for very few electrons in the island.


\section{Conclusions}

In this paper we have shown that it is possible to turn electrons into
anyons if they are placed under an elastic membrane that is both
charged and magnetized. We have also proposed an experimental set up
to detect the anyonic statistics, and estimated the size of the
measurable phases for a possible concrete experimental realization
using patterned metallic rings and islands under a suspended graphene
sheet, which could be both charged and doped with magnetic impurities.

Finally, let us point out that there is nothing intrinsic to our
general ``membrane'' mechanism which necessitates that the
``electronyons'' feel an induced logarithmic potential. Indeed, had we
not charged the membrane at all, the electrons would still become
anyons due to their Zeeman coupling to the magnetic field (albeit for
distances less than $L_\tau$, which could be made large by not
tensioning the membrane). However, there would not be a logarithmic
potential, as the induced coupling would be of the form
$\nabla^2\phi~\psi^\dagger\sigma_z \psi$. The resulting statistical
parameter would be too small in this case.

One could imagine yet a different setup: instead of one membrane,
there are three membranes on top of each other, with charge densities
$\sigma_e$, $-2\sigma_e$, and $\sigma_e$, for the top, middle, and
bottom membrane respectively. As long as the membranes are very close to
one other, and can be made to move together, there will be an
effective quadrpole charge density.  This will result in a coupling of
the form $\nabla^2\phi~\psi^\dagger \psi$, which again does not induce
an effective force between electrons.  In this scenario, ignoring the
Coulomb force between electrons, the theory will be precisely that of
\cite{Katz} at distances less than $L_\tau$.

\appendix 

\section{The electric and vector potentials due to membrane
  fluctuations}

In this Appendix we present the potentials $(A_0^\phi,\vec{A^\phi})$
due to the membrane phonon. We will present the vector potential
calculation in detail. The electric potential is found in a very
similar manner. We assign the $X'=(x',y')$ coordinate to the membrane
and $X=(x,y)$ to the fixed plane. We are interested in the
contribution from $\phi(x,y)$. The vector potential on the fixed plane
is calculated from:
    

\begin{widetext}
\begin{eqnarray}
\label{Coulomb}
A_i(X)
&=&
-\epsilon_{ij}\nabla_j \int\,d^2 X'\,\frac{\sigma_m}{\left(\left(X'-X\right)^2+\left(\delta+\phi\left(X'\right)\right)^2\right)^{1/2}}
\;\;
\nonumber\\
&=&
-\frac{4\pi}{\left(2\pi\right)^3}\sigma_m\,\epsilon_{ij}\nabla_j \int d^2 X'\,\int d^2 K'\,\int d \kappa
\;\;\frac{e^{i{\vec{K'}.\left(\vec{X}-\vec{X'}\right)}}
\;e^{i\kappa\left(\delta+\phi(X')\right)}}{K'^2\,+\,\kappa'^2}
\nonumber\\
&=&
-\frac{4\pi}{\left(2\pi\right)^2}\sigma_m\,\epsilon_{ij}\nabla_j \int d^2 X'\,\int d^2 K'\;\;
\frac{e^{i{\vec{K'}.\left(\vec{X}-\vec{X'}\right)}}
\;e^{-\left|K'\right|\left|\delta+\phi(X')\right|}}{2\left|K'\right|}
\;.
\end{eqnarray}
The $\phi$-dependence of the vector potential can be extracted by
linearizing in $\phi$, in the regime that the membrane does not touch
the 2DEG, when ${\rm sgn} (\delta+\phi)={\rm sgn}(\delta)$:
\begin{eqnarray}
A^\phi_i(X)
&=&
\frac{2\pi}{\left(2\pi\right)^2}\sigma_m\,\epsilon_{ij}\nabla_j \int d^2 X'\,\int d^2 K'\;\;
e^{i{\vec{K'}.\left(\vec{X}-\vec{X'}\right)}}
\;e^{-\left|K'\right||\delta|}
\;\phi(X')\;{\rm sgn}(\delta)
\;.
\end{eqnarray}
We are interested in small fluctuations of the field and in momenta
$|K'|\,\delta\ll1$.  Consequently,
\begin{eqnarray}
A^\phi_i(X)
&\approx&
2\pi\sigma_m \,{\rm sgn}(\delta)
\;\epsilon_{ij}\,\nabla_j\left(\phi(X)\,-\,\int\frac{d^2K'}{(2\pi)^2}\,\left|K'\right||\delta|
\;e^{i\vec{K'}.\vec{X}}
\;\tilde{\phi}(K')\right)
\;.
\end{eqnarray}
\end{widetext}
Here $\tilde\phi$ is the Fourier transform of $\phi$, and we have
included the leading correction in $\delta$.  It is interesting to
note that corrections are non-local, furthermore we see that $\delta$
appears with factors of $K'$, therefore the corrections decouple in the
long wavelength limit.  An analogous formula holds for the electric
potential, $A_0^\phi(X)$, with $\sigma_m$ replaced by $\sigma_e$ (and
with the $-\epsilon_{ij} \nabla_j$ derivative absent).

\acknowledgments

We thank Maissam Barkeshli, Alex Kitt, and Yiming Xu for useful
discussions.  This work was supported in part by DOE grant
DEFG02-01ER-40676 and NSF CAREER grant PHY-0645456 (SRB and EK) and
DOE Grant DEFG02-06ER46316 (CC).


\end{document}